\documentclass[fleqn,twoside]{article}
\usepackage[headings]{espcrc2}

\readRCS
$Id: espcrc2.tex,v 1.2 2004/02/24 11:22:11 spepping Exp $
\usepackage{graphicx}
\usepackage[figuresright]{rotating}

\graphicspath{{./}{./figures/}}

\def\bbuildrel#1_#2^#3%
{\mathrel{\mathop{\kern 0pt#1}\limits_{#2}^{#3}}}

\newcommand{\ovl}[1]{\overline{#1}}
\newcommand{\ice}[1]{\relax}
\newcommand{\re}[1]{(\ref{#1})}

\newcommand{\beq}{\begin{equation}}
\newcommand{\eeq}{\end{equation}}
\newcommand{\bea}{\begin{eqnarray}}
\newcommand{\eea}{\end{eqnarray}}

\newcommand{\ba}{\begin{array}}
\newcommand{\ea}{\end{array}}

\newcommand{\als}{\alpha_s}

\newcommand{\dsp}{\displaystyle}

\newcommand{\smsp}{  }
\newcommand{\sbz}{  }

\newcommand{\dFabcFabcdRinvAnlA}{n_f\smsp \frac{d^{abc}\smsp d^{abc}}{d_R}}
\newcommand{\dFabcFabcdRinvA}{\smsp \frac{d^{abc}\smsp d^{abc}}{d_R}}

\let\*=\,

\newcommand{\dd}{{\rm d}}
\newcommand{\nn}{\nonumber}
\newcommand{\nnb}{\nonumber}

\newcommand{\AmS}{{\protect\the\textfont2
  A\kern-.1667em\lower.5ex\hbox{M}\kern-.125emS}}

\title{
{
 \vspace*{-14mm}
\centerline{\normalsize\hfill TTP10-31}
\baselineskip 11pt
{}}
\vspace{3mm}
\vspace{1cm}
Adler Function, DIS sum rules and Crewther Relations
\thanks{
Talk presented at 10-th DESY Workshop on Elementary Particle Theory:
Loops and Legs in Quantum Field Theory, W\"orlitz, Germany, 25-30 April  2010.
}}

\author{P.~A.~Baikov\address{Skobeltsyn Institute of Nuclear Physics,
        Moscow State University, \\
        Moscow~~119991, Russia}%
       ,
        K.G.Chetyrkin\address[KUNI]{
Institut f\"ur Theoretische Teilchenphysik,
  KIT, D-76128 Karlsruhe, Germany
        }%
\thanks{On leave from Institute for Nuclear Research
of the Russian Academy of Sciences, Moscow, 117312, Russia.} 
            and 
J.~H.~K\"uhn\addressmark[KUNI]
}

\runtitle{2-column format camera-ready paper in \LaTeX}
\runauthor{K.G. Chetyrkin}

\begin{document}

\begin{abstract}

The current status of the Adler function and two closely
related Deep Inelastic Scattering (DIS) sum rules, namely,  the
Bjorken sum rule for polarized DIS and the Gross-Llewellyn Smith sum
rule are briefly reviewed.  A new result is  presented: an analytical calculation of the
coefficient function of the latter  sum rule in a generic gauge theory in order
${\cal O}(\als^4)$. It is demonstrated that the corresponding
Crewther relation allows to fix two of  three colour structures 
in the ${\cal O}(\als^4)$ contribution to the singlet 
part of the Adler function.  


\vspace{1pc}
\end{abstract}

\maketitle

\section{Introduction}

Exactly ten years ago at ``Loops and Legs 2000'' \cite{Chetyrkin:2000uw}
one of the present authors  discussed  the  
perspectives  of computing the famous ratio  
$
  R(s) = {\sigma(e^+e^-\to {\rm hadrons})\over \sigma(e^+e^-\to
    \mu^+\mu^-)}\,
$
or, equivalently,  the Adler function  of the correlator  of the EM vector currents
at order $\als^4$ in massless QCD (for a general review see \cite{Chetyrkin:1996ia}).
The main conclusion was
that
`` 
\dots a better understanding
of all kinds of  relations connecting various p-integrals
could eventually result to the reduction of an arbitrary 5-loop
p-integral to a combination of some limited number (a few dozens?) of
master p-integrals\footnote{By  ``p-integrals'' we understand   massless propagator-like (that
is  dependent on only one external momentum) Feynman integrals.}.  Once it is done
it  should be not very difficult to evaluate the latter analytically
or numerically \dots'' 
Now, ten years later, we can rightfully
state that the program has been successfully 
worked out and almost (see below) completed. 
\cite{Baikov:2001aa,Baikov:2005rw,Baikov:2008jh,PhysRevLett.104.132004,Baikov:2010hf,Smirnov:2010hd}.

\ice{
Even more, the use of 
the (generalized) Crewther relations

Indeed, let us list main steps
which  have been done. 
\item{} z
\item{}
}

Below we summarize the current status of the calculations of the Adler
function and of  two other, through generalized
Crewther relations \cite{Crewther:1972kn,Broadhurst:1993ru}  closely related, physical observables: the Bjorken and the
Gross-Llewellyn Smith DIS sum rules \cite{Bjorken:1967px,Gross:1969jf} at order $\als^4$ in
massless QCD.


\section{Adler Function}

It is convenient to start with the polarization  operator of the flavor singlet vector current:
\beq
3\smsp  Q^2\smsp  \Pi(Q^2) =
i\int\dd^4 xe^{iq\cdot x}\langle 0|{\rm  T}j_\mu(x)j^\mu(0)|0\rangle
{},
\eeq
with $j_{\mu} = \sum_i\ovl{\psi}_i\gamma_{\mu}\psi_i$ and  $Q^2 = -q^2$.
The  corresponding Adler function
\beq
{D}(Q^2) =  -12\, \pi^2
Q^2\, \frac{\mathrm{d}}{\mathrm{d} Q^2} \Pi(Q^2)
{}
\eeq
is naturally decomposed into a sum of the nonsiglet (NS) and singlet (SI) components (see Fig.~1):
\beq
{D}(Q^2) = n_f\, \,  D^{NS}(Q^2) + n_f^2 \, D^{SI}(Q^2)
\label{D:decomp}
 {}.
\eeq
Here 
$n_f$ stands for   the total number of quark flavours; all quarks  are  considered as massless.

\begin{figure*}
\begin{center}
\includegraphics[width=12cm]{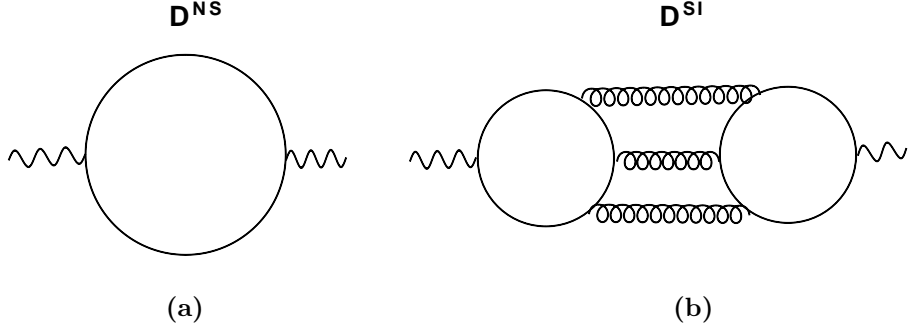}
\end{center}

\hspace{4cm} {\bf (a)}  \hspace{6cm} {\bf (b)}

\caption{
(a),(b): lowest order nonsinglet and singlet diagrams contributing to the Adler function.
}
\label{table:1}
\end{figure*}

Note that  the Adler function $D^{EM}$ corresponding to the electromagnetic vector current
$j^{EM}_{\mu} = \sum_i Q_i \ovl{\psi}_i\gamma_{\mu}\psi_i$ ($Q_i$ stands for the electric charge of
the quark field $\psi_i$)  is given by the expression:
\beq
{D}^{EM} = \left(  \sum_i Q_i^2 \right) D^{NS} + 
\left(  \sum_i Q_i \right)^2
 \, D^{SI}
 {}.
\eeq

The perturbative expansions of  nonsinglet and singlet 
parts read ($a_s \equiv \frac{\alpha_s}{\pi}$):
\bea
{D}^{NS}(Q^2) &=& d_R \left( 
1+
\sum_{i=1}^{\infty} \  {d}^{NS}_i \, a_s^i(Q^2)
\right)
{},
\\
{D}^{SI}(Q^2) &= & d_R \left(
\sum_{i=3}^{\infty} \  {d}^{SI}_i \, a_s^i(Q^2)
{}
\right)
{},
\eea
where for future convenience the parameter   $d_R$ (the dimension of the quark color representation, $d_R= 3$ in QCD)
is  factorized in {\em both} nonsinglet and singlet components. 

At order $\als^3$ both components  of the Adler function are known since long 
\cite{Gorishnii:1990vf,Surguladze:1990tg,Chetyrkin:1996ez} for the  case of  
a general  colour   gauge group. The corresponding   $\als^4$ calculation has been recently  finished
\cite{Baikov:2008jh,PhysRevLett.104.132004}
for the nonsinglet function ${D}^{NS}$ {\em only} .

The singlet component has the following structure at orders $\als^3$ and $\als^4$:
\beq
{d}^{SI}_3 
= \dFabcFabcdRinvA \left(
\frac{11}{192}
-\frac{1}{8}  \sbz \zeta_{3}
\right)
{},
\eeq
\bea 
\lefteqn{d_4^{SI} =}
\label{d4SI}
\\
&{}&
 \dFabcFabcdRinvA \left(
C_F \, d_{4,1}^{SI} +C_A \, d_{4,2}^{SI} + T\, n_f\, d_{4,3}^{SI}
\right)
{}
\nn
{}.
\eea
Here  $C_F$ and $ C_A$ are the quadratic Casimir
operators of the
fundamental and the adjoint representation of the Lie algebra,
$d^{abc} = 2\, \mathrm{Tr}(\{\frac{\lambda^a}{2}\frac{\lambda^b}{2}\}  \frac{\lambda^c}{2}\}$,
$T$ is the trace normalization of the
fundamental representation. 
\ice{
The exact definitions of
${d_F^{a b c } d_F^{a b c }}$
are given in \cite{Vermaseren:1997fq}.
}
For  QCD (colour gauge group SU(3)):
\[
C_F =4/3\,,\, C_A=3\,,\, T=1/2\,,
{d^{a b c } d^{a b c }} = \frac{40}{3}
\nonumber
{}.
\]

From general considerations we expect that the missing $\als^4$
contribution to the singlet component of the Adler function should be
numerically inessential  (see the next section for  an explicit argument in
favour of this assumption). The corresponding direct calculation is under way
and should be finished in the   near future.

\ice{
Let consider in  f

For future reference  we give below 

Nevertheless, already
Some explicit ar
results 
The current status of the both Adler functions is as follows.
The calculation of 
}

\section{ DIS sum rules}
The Bjorken sum rule
expresses
the integral over the  spin distributions of quarks inside of the nucleon in terms of
its axial charge times a  coefficient function
${C}^{Bjp}$:
\bea
\Gamma_1^{p-n}(Q^2) &=&
\int_0^1 [g_1^{ep}(x,Q^2)-g_1^{en}(x,Q^2)]dx
\nnb
\\
&=&\frac{g_A}{6}
C^{Bjp}(a_s) +
\sum_{i=2}^{\infty}\frac{\mu_{2i}^{p-n}(Q^2)}{Q^{2i-2}}
{},
\label{gBSR}
\eea
where $g_1^{ep}$ and $g_1^{en}$ are the spin-dependent proton and neutron
structure functions, $g_A$ is the nucleon axial charge as measured in
neutron $\beta$-decay. The coefficient function
$C^{Bjp}(a_s)=1+{\cal O}(a_s)$  is proportional to the flavour-nonsinglet axial vector
current $\bar{\psi}\gamma^{\mu}\gamma_5 t^a\psi$ in the corresponding
short distance Wilson expansion. The sum  in the second line of \re{gBSR} describes 
the nonperturbative  power corrections (higher twist) which are  inaccessible for pQCD.

\begin{figure*}
\begin{center}
\includegraphics[width=12cm]{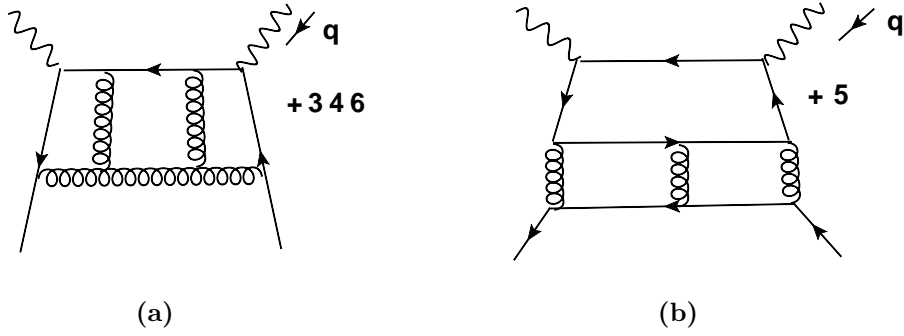}
\end{center}
\hspace{3.6cm} {\bf (a)}  \hspace{6.2cm} {\bf (b)}
\caption{
(a),(b): ${\cal O}(\als^3)$ nonsinglet and singlet diagrams contributing to the Gross-Llewellyn Smith sum rule;
note that the  coefficient function  $C^{Bjp}$ is contributed by only nonsinglet diagrams.
}
\label{table:1}
\end{figure*}

Another, closely related sum rule, the Gross-Llewellyn Smith one, reads
(we do not write explicitly the higher twist corrections below) 
\begin{equation}
\frac{1}{2}\int_0^{1} F_3^{\nu p+\overline{\nu}p}(x,Q^2)dx= 3\, C^{GLS}(a_s)
\nnb
{},
\end{equation}
where $ F_3^{\nu p+\overline{\nu}p}(x,Q^2)$ is the isospin singlet structure function. 
The function $C^{GLS}(a_s)$ comes from
operator-product expansion of the axial and vector  nonsinglet
currents
\begin{equation}
i\int T{A_{\mu}^{a}(x)V_{\nu}^{b}(0)}e^{iqx}dx|_{q^2\rightarrow{\infty}}
\approx C_{\mu\nu\alpha}^{V,ab}\,V_{\alpha}(0)+ \dots
\nnb
\end{equation}
where 
\[C_{\mu\nu\alpha}^{V,ab}\sim
i\delta^{ab}\epsilon_{\mu\nu\alpha\beta}
\frac{q^{\beta}}{Q^2}C^{GLS}(a_s)
{}
\]
and
$V_{\alpha} =\ovl{\psi}\gamma_{\alpha} \psi$ is a flavour singlet  quark current.
At last 
$A_{\mu}^{a} = \ovl{\psi}\gamma_{\mu} \gamma_5 t^a \psi$,  
$V_{\nu}^{b} = \ovl{\psi}\gamma_{\nu} t^b \psi$ are axial vector    and  vector 
nonsinglet quark currents, with the $t^a, \, t^b$  being being the generators of the flavour group
$SU(n_f)$.

All diagrams contributing to $C^{GLS}(a_s)$ can be separated in two groups:
nonsinglet and singlet ones (see Fig. 2):
\bea
C^{GLS} &=& C^{NS} + C^{SI},
\label{GLSdecomp}
\\
C^{NS}(Q^2) &=& 
1+
\sum_{i=1}^{\infty} \  c^{NS}_i \, a_s^i(Q^2)
{},
\\
{C}^{SI}(Q^2) &= &
\sum_{i=3}^{\infty} \  c^{SI}_i \, a_s^i(Q^2)
{}.
\eea

The results for both functions  $C^{Bjp}$ and $C^{GLS}$ at order 
$\als^3$ are known since early 90-ties \cite{Larin:1991tj}. 
Note that there is  a remarkable connection  (following from the chiral invariance \cite{Larin:1991tj})
\beq
C^{Bjp} \equiv C^{NS}
\label{chiral:inv}
\eeq
valid in all orders of the perturbation theory.

The ${\cal O}(\als^4)$ contribution to $C^{Bjp}$  has been recently computed  and
published in \cite{PhysRevLett.104.132004}. 
The calculation of the  ${\cal O}(\als^4)$ contribution to  ${C}^{SI}$
has been just finished. The results for both functions are given  below.

We start from their  numerical form. 
\begin{eqnarray}
&{}& C^{NS} =
1
{-} a_s
{+}
\left(
-4.583
+0.3333  \,n_f
\right)a_s^2
\\
&{+}& a_s^3
\left(
-41.44
+7.607  \,n_f
-0.1775  \, n_f^2
\right) a_s^3
\nonumber\\
&{+}&
\left(
-479.4
+123.4  \,n_f
-7.697  \, n_f^2
+0.104  \, n_f^3
\right)\,a_s^4
{},
\nonumber
\label{CBJN}
\end{eqnarray}
\beq
{}C^{SI}  =  0.4132\,n_f\,  a_s^3 +a_s^4\, n_f
\left(
{
5.802 -  0.2332\,n_f
}
\right)
{}.
\label{SI}
\eeq
For a typical value of $n_f=3$ the  above relations read:
\begin{eqnarray}
{} C^{NS}(n_f=3) &=&
1
-1.  a_s
-3.583  a_s^2
\\
&{-}&
20.22  a_s^3
 -175.7  \,a_s^4
\nonumber
\label{CBJNnf3}
{},
\end{eqnarray}
\beq
{}C^{SI}  = 1.2396\,  a_s^3 + 15.3072\,  a_s^4
{}.
\eeq

\ice{

r=  Expand[0.4132*nf* as^3 +as^4*nf*(5.802 -  0.2332*nf)]/.nf->3

In[3]:= r=  Expand[0.4132*nf* as^3 +as^4*nf*(5.802 -  0.2332*nf)]/.nf->3
                 3             4
Out[3]= 1.2396 as  + 15.3072 as
}

As expected the singlet contributions are  less than the nonsinglet ones
by at least one order in magnitude for  each from two available orders in $\als$.

The result for $C^{NS}$ valid for  a generic gauge group is given in 
\cite{PhysRevLett.104.132004}. For the singlet coefficient function 
the generalization of eq.~(\ref{SI}) to a  generic gauge group read

\beq c_3^{SI} = \dFabcFabcdRinvAnlA \left(
c_{3,1}^{SI} \equiv
-\frac{11}{192}
+\frac{1}{8}  \sbz \zeta_{3}
\right)
{},
\eeq
\bea 
\lefteqn{c_4^{SI} =}
\label{c4SI}
\\
&{}&
 \dFabcFabcdRinvAnlA \left(
C_F \,c_{4,1}^{SI} +C_A \, c_{4,2}^{SI} + T\, n_f\, c_{4,3}^{SI}
\right)
{},
\nn
\eea
\beq
 c_{4,1}^{SI} = {
\frac{37}{128}
+\frac{1}{16}  \sbz \zeta_{3}
-\frac{5}{8}  \sbz \zeta_{5}
}
\label{c41SI}
{},
\eeq
\beq
c_{4,2}^{SI} =  { -\frac{481}{1152}
+\frac{971}{1152}  \sbz \zeta_{3}
-\frac{295}{576}  \sbz \zeta_{5}
+\frac{11}{32}  \sbz \zeta_3^2
}
{},
\label{c42SI}
\eeq

\beq
c_{4,3}^{SI} = {
\frac{119}{1152}
-\frac{67}{288}  \sbz \zeta_{3}
+\frac{35}{144}  \sbz \zeta_{5}
-\frac{1}{8}  \sbz \zeta_3^2
}
\label{c43SI}
{}.
\eeq

\ice{
The polarized Bjorken sum rule  reads
\beq
Bjp(Q^2)=\int_0^1[g_1^{ep}(x,Q^2)-g_1^{en}(x,Q^2)]dx=\frac{1}{6}
|\frac{g_A}{g_V}|C^{Bjp}(a_s)
\eeq
Coefficient function $C^{Bjp}(a_s)$ is  fixed by OPE
of two nonsinglet vector
currents (up to power suppressed corrections)
\begin{equation}
i\int T{V_{\alpha}^{a}(x)V_{\beta}^{b}(0)}e^{iqx}dx|_{q^2\rightarrow{\infty}}
\approx C_{\alpha\beta\rho}^{Q,abc}A_{\rho}^{c}(0)+\dots
\label{7}
\end{equation}
where
\begin{equation}
C_{\alpha\beta\rho}^{Q,abc}\sim
id^{abc}\epsilon_{\alpha\beta\rho\sigma}
\frac{q^{\sigma}}{Q^2}C^{Bjp}(a_s)
\nnb
\end{equation}
and $Q^2=-q^2$
}

\ice{
\section{Crewther relations and DIS sum rules}

The Crewther relation relates in a 

}

\section{Crewther relations}

There exist two  (generalized) Crewther relations which  connect the nonsinglet 
and the full Adler functions to the coefficient functions $C^{Bjp}$ and $C^{GLS}$ respectively
\cite{Broadhurst:1993ru}. The relations state that
\bea
&{}& \hspace{-.9cm} D^{NS}(a_s)\, C^{Bjp}(a_s) = d_R \left[ 1+ \frac{\beta(a_s)}{a_s}\, K^{NS}
\right]
\label{gCrewtherNS}
\, ,
\\
&{}& K^{NS}\equiv K^{NS}(a_s) =
a_s\,K^{NS}_1 
\nn
\\
&& \hspace{1.7cm} + a_s^2\,K^{NS}_2 +a_s^3\,K^{NS}_3
+ \dots
\nn
\eea
and 
\bea
&{}& \frac{ \dsp D(a_s)\,  C^{GLS}(a_s)}{d_R \, n_f}
\label{gCrewtherFull}
= 
 1+ 
\frac{\beta(a_s)}{a_s}\, K(a_s),
\\
&{}& K(a_s) =
a_s\,K_1 
\nn
\\
&{}& \hspace{1.7cm} + a_s^2\,K_2 +a_s^3\,K_3
+ \dots
\nn
\eea
Here $\beta(a_s) = \mu^2\,\frac{\mathrm{d}}{\mathrm{d} \mu^2}
a_s(\mu) =-\sum_{i \ge 0} \beta_i a_s^{i+2}$  is the QCD
$\beta$-function with its first term $ \beta_0 = \frac{11}{12}\, C_A -\frac{T}{3}\,n_f$. 
The term proportional to the $\beta$-function describes the deviation from
the limit of exact conformal invariance, with the deviations starting in order $\alpha_s^2$.
The relations  \re{gCrewtherNS} and \re{gCrewtherFull} were suggested in \cite{Broadhurst:1993ru} on the basis of
 \cite{Crewther:1972kn,Adler:1973kz} and the 
${\cal O}(\alpha_s^3)$ calculations of the functions $C^{Bjp}, \, C^{GLS}$ and $D$
carried out in\cite{Gorishnii:1990vf,Surguladze:1990tg,Larin:1991tj}. 
Formal proofs were considered in \cite{Crewther:1997ux,Braun:2003rp}.


Relation \re{gCrewtherNS} has been investigated in work \cite{PhysRevLett.104.132004}.  Here it was demonstrated 
that  at orders $\alpha_s^2, \alpha_s^3$ and $\alpha_s^4$ 
eq.~\re{gCrewtherNS} produces  as many as 2, 3 and, finally, 6 constraints on the
combinations $d^{NS}_2 + c^{NS}_2, d^{NS}_3 +c^{NS}_3$ and $d^{NS}_4 +c^{NS}_4$ respectively
(for a very detailed  discussion at orders $\alpha_s^2 $ and $\alpha_s^3$
see  also  \cite{Broadhurst:1993ru}).

The fulfillment of these constraints has provided us with a  powerful
test of the correctness of the calculations of $D^{NS}(a_s)$ and
$C^{Bjp}(a_s)$. It also fixes unambiguously the (nonsinglet) Crewther
parameters $K^{NS}_1,K^{NS}_2$ ans $K^{NS}_3$ (for explicit expressions see \cite{PhysRevLett.104.132004}).

Let us consider now eq.~\re{gCrewtherFull} (assuming that
\re{gCrewtherNS} is fulfilled). Combining  
eqs.~(\ref{D:decomp},\ref{GLSdecomp},\ref{chiral:inv}) and \re{gCrewtherNS} leads  
to the following relations between coefficients $K_i^{NS}$ and $K_i$:
\bea
K_1 &=& K_1^{NS}, \ \ K_2 = K_2^{NS},
\\
 K_3 &=& K^{NS}_3 + K_3^{SI},
\\
K_3^{SI} &=& k^{SI}_{3,1} \,\, \dFabcFabcdRinvAnlA
{}.
\eea

Thus, we conclude that eq. ~\re{gCrewtherFull} puts $3-1 = 2$
constraints between two triplets of (purely numerical) parameters $\{
d^{SI}_{4,1}, d^{SI}_{4,2}, d^{SI}_{4,3}\}$ and $\{ c^{SI}_{4,1},
c^{SI}_{4,2}, c^{SI}_{4,3}\}$ appearing in eqs.~\re{d4SI} and \re{c4SI}) and
completely describing the order $\alpha_s^4$ singlet contributions to
the Adler function and the Gross-Llewellyn Smith sum rule
respectively.

The solution of the constraints and eqs.~(\ref{c41SI}-\ref{c43SI}) 
produces  the following  result for $d_4^{SI}$:
\bea d_{4,1}^{SI} &=& -\frac{3}{2}c_{3,1}^{SI} - c_{4,1}^{SI}
 =
 -\frac{13}{64} - \frac{\zeta_3}{4}  + \frac{5\zeta_5}{8} 
{},
\eea
\bea
d_{4,2}^{SI} &=& - c_{4,2}^{SI}  -  \frac{11}{12}\, k_{3,1}^{SI}
{},
\eea
\bea
 d_{4,3}^{SI} &=& - c_{4,3}^{SI}  +  \frac{1}{3} \,k_{3,1}^{SI} 
{}.
\eea

\ice{
\bea
&{}& \left(D^{NS} + \frac{n_f}{d_R}\,D^{SI}\right) C^{GLS} 
=
\\
&{}&
 1+ \frac{\beta(a_s)}{a_s}\, K(a_s),
\\
&{}& K^{NS}(a_s) =
K^{NS}_0 +
a_s\,K^{NS}_1 
\nn
\\
&& \hspace{1.7cm} + a_s^2\,K^{NS}_2 +a_s^3\,K^{NS}_3
+ \dots
\label{gCrewtherFull}
\eea
}

\section{Conclusion}

We have analytically computed the ${\cal O}(\als^4)$ contribution to
the Gross-Llewellyn Smith sum rule. The result taken together with the
corresponding (generalized) Crewther relation leads to a prediction for 
 the (still unknown) ${\cal O}(\als^4)$ term in the singlet component of
the Adler function.  The prediction depends on only one unknown 
numerical parameter --- the Crewther coefficient $k^{SI}_{3,1}$.

The direct calculation of  the three coefficients 
$\{d^{SI}_{4,1}, d^{SI}_{4,2}, d^{SI}_{4,3}\}$ parameterizing the ${\cal O}(\als^4)$ 
contribution to the singlet component of the Adler function will  be finished soon. 
Then  will get another strong check of the  complicated machinery
employed in performing the calculations.

The calculation of the coefficient function $C^{GLS}$ has been performed
on a SGI ALTIX 24-node IB-interconnected cluster of 8-cores Xeon
computers and on the HP XC4000 supercomputer of the federal state
Baden-W\"urttemberg
using  parallel  MPI-based \cite{Tentyukov:2004hz} as well as thread-based 
\cite{Tentyukov:2007mu} versions  of FORM
\cite{Vermaseren:2000nd}.  For the evaluation of color factors we have used the FORM program {\em COLOR}
\cite{vanRitbergen:1998pn}. The diagrams have been generated with QGRAF \cite{Nogueira:1991ex}.
The figures have been drawn with the  the help of
Axodraw \cite{Vermaseren:1994je} and JaxoDraw  \cite{Binosi:2003yf}.

This work was supported by the Deutsche Forschungsgemeinschaft in the
Sonderforschungsbereich/Transregio SFB/TR-9 ``Computational Particle
Physics'' and  by RFBR grants 08-02-01451, 10-02-00525.

\end{document}